\documentclass[aps,prb,twocolumn,amsmath,amssymb,superscriptaddress]{revtex4}
\usepackage{graphicx}
\usepackage{amssymb}
\usepackage{natbib}
\usepackage{color}

\UseRawInputEncoding

\newcommand{\beq}{\begin{eqnarray}}
\newcommand{\eeq}{\end{eqnarray}}

\newcommand{\RomanNumeralCaps}[1]
    {\MakeUppercase{\romannumeral #1}}

\begin{document}

\title{Topological Hall effect and the magnetic states of Nowotney chimney ladder compound Cr$_{11}$Ge$_{19}$}
\author{Yu Li}
\email{yuli1@lsu.edu}
\affiliation{Department of Physics $\&$ Astronomy, Louisiana State University, Baton Rouge, LA 70803, USA}

\author{Xin Gui}
\affiliation{Department of Chemistry, Louisiana State University, Baton Rouge, Louisiana 70803, USA}

\author{Mojammel A Khan}
\affiliation{Department of Physics $\&$ Astronomy, Louisiana State University, Baton Rouge, LA 70803, USA}
\affiliation{Department of Physics and Astronomy $\&$ Department of Chemistry, Johns Hopkins University, Baltimore, MD 21210, USA}

\author{Weiwei Xie}
\affiliation{Department of Chemistry, Louisiana State University, Baton Rouge, Louisiana 70803, USA}

\author{David P. Young}
\affiliation{Department of Physics $\&$ Astronomy, Louisiana State University, Baton Rouge, LA 70803, USA}

\author{J. F. DiTusa}
\email{ditusa@phys.lsu.edu}
\affiliation{Department of Physics $\&$ Astronomy, Louisiana State University, Baton Rouge, LA 70803, USA}

\date{\today}

\begin{abstract}
We have investigated the magnetic and charge transport properties of single
crystals of Nowotney Chimney Ladder compound Cr$_{11}$Ge$_{19}$ and
mapped out a comprehensive phase diagram reflecting the complicated
interplay between the Dzyaloshinskii-Moriya (DM) interaction, the
dipolar interaction, and the magnetic anisotropy. We have identified a
set of interesting magnetic phases and attributed a finite topological Hall effect to the recently discovered bi-skyrmion phase. These data also suggest the existence of an anti-skyrmion state at finite fields for temperatures just below the
magnetic ordering temperature, $T_c$, as indicated by a distinct change in sign of the topological Hall effect. Above $T_c$, we discovered a region of enhanced magnetic response corresponding to a disordered phase likely existing near the ferromagnetic critical point under small magnetic fields. Strong spin chirality fluctuations are demonstrated by the large value of the topological Hall resistivity persisting up to 1 T which is most likely due to the existence of the DM interaction. We argue that changes to the topological Hall effect correspond to different topological spin textures that are controlled by magnetic dipolar and
DM interactions that vary in importance with temperature.
\end{abstract}

\maketitle

\section{INTRODUCTION}
Skyrmions and anti-skyrmions are nanoscale particle-like spin textures
found in a variety of systems from chiral structured magnets to thin
ferromagnetic films \cite{Skyrmion1,Skyrmion2,Skyrmion3}. Each
skyrmion or anti-skyrmion carries a positive or negative topological
charge known as the scalar spin chirality. When a skyrmion meets an
anti-skyrmion, they are expected to annihilate and emit
magnons\cite{Dynamics}, in analogy to the annihilation of matter and
antimatter generating electromagnetic radiation. These topological magnetic spin textures possess a net scalar spin chirality which is implicitly
determined by the crystal structure and dominant magnetic
interactions. For instance, skyrmion phases in chiral magnets are
caused by the isotropic Dzyaloshinskii-Moriya (DM)
interaction\cite{Nagaosa2013,DM_DiTusa} while anti-skyrmions found in
some Heusler alloys are due to an anisotropic DM interaction
with opposite signs along the $x$ and $y$
axes\cite{Nayak2017,Huang2017} that is favored by the $D_{2d}$ crystal
symmetry. Moreover, in many centrosymmetric magnets in the absence of
the DM interaction, the interplay between the dipolar interaction and
the magnetic anisotropy leads to the formation of bi-skyrmions, a bound pair of skyrmions with opposite
helicity\cite{Biskyrmion1,Biskyrmion2,Biskyrmion3,Biskyrmion4}. In
principle, one could tune the anisotropic DM interaction with respect
to the dipolar energies and magnetic anisotropies to create a
transition between skyrmion states and anti-skyrmion states.
Recently£¬ it was reported\cite{transition1,transition2} that a topological transformation from antiskyrmions
to skyrmions occurs in noncentrosymmetric Mn$_{1.4}$Pt$_{0.9}$Pd$_{0.1}$Sn, making
the $D_{2d}$ systems especially interesting.

\begin{figure}[htb]
\includegraphics[scale=2]{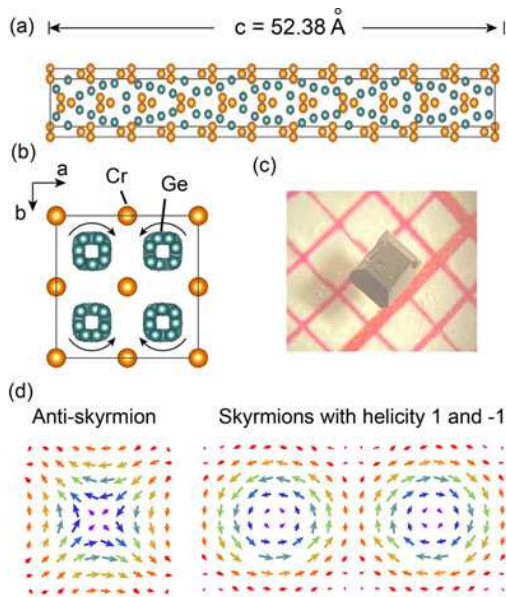}
\caption{ Crystal structure, morphology, and skyrmion schematics. Crystal structure of Cr$_{11}$Ge$_{19}$ (a) along the $c$-axis and (b) in the $ab$ plane. Orange (blue) spheres are Cr(Ge) atoms. The arrows represent the direction of rotation of the Ge
helices. The surrounding Cr helices have the opposite sense of
rotation. (c) Picture of a Cr$_{11}$Ge$_{19}$ crystal. The box size on the graph paper is 1$\times$1 mm$^2$. (d)
In-plane spin configuration of an anti-skyrmion and a bi-skyrmion
pair. The red color represents spin up while purple color corresponds to
spin down. }
\label{Fig1}
\end{figure}

Among the few topological magnetic materials known, Cr$_{11}$Ge$_{19}$
is unique as it is the first non-centrosymmetric compound found to
host a bi-skyrmion spin texture\cite{Takagi2018}. Its crystal
structure contains 4-fold helices of Cr atoms along the c-axis with
separate helices of Ge nested inside [Fig.~\ref{Fig1}(a) and
(b)]. Cr$_{11}$Ge$_{19}$
is one of the two known member of the Nowotney Chimney Ladder
compounds\cite{NCladder,Mn3Ge5} to display a magnetically ordered ground
state. Its crystal structure has the $D_{2d}$ symmetry thought to support an
antisymmetric DM interaction favoring anti-skyrmion formation, as is
the case in the tetragonal Heusler compound
Mn$_{1.4}$Pt$_{0.9}$Pd$_{0.1}$Sn\cite{Nayak2017}.  In contrast to the expectation of anti-skyrmions, only
bi-skyrmions were reported in a recent Lorentz transmission electron
microscopy (LTEM) experiment performed on a thin lamella at 6 K and a field of 40 mT,
suggesting the dominance of dipolar
interactions\cite{Takagi2018}. Generally, the magnetic dipolar
interaction is long-ranged and can be screened by spin fluctuations as
the system approaches the magnetic transition temperature while the DM
interaction is thought to be temperature independent. This provides an opportunity to tune the strength of the relevant interactions
with temperature and sample geometry in anticipation of a topological phase of
anti-skyrmions in Cr$_{11}$Ge$_{19}$ when the DM interaction is
dominant.

In this manuscript, we report an extensive investigations of the
magnetic and charge transport properties of high quality
Cr$_{11}$Ge$_{19}$ single crystals. We have mapped out a comprehensive
phase diagram and indicate a set of interesting magnetic phases in
Cr$_{11}$Ge$_{19}$ that are dependent on temperature and field. These
data suggest that the recently discovered bi-skyrmion phase\cite{Takagi2018}
of this material is just one of several interesting magnetic
phases. A tendency for the redistribution of spin textures towards the stripe ordered state is suggested
at low temperatures while a strongly fluctuating phase (or region) at small magnetic fields and above the magnetic ordering temperature
is indicated by the magnetic susceptibility. Based upon measurements of the topological Hall effect (THE),
we speculate that an anti-skyrmion phase exists at finite fields and
temperatures approaching the magnetic ordering temperature $T_c$ that
is highly favored by the crystalline structure.

\section{SAMPLE PREPARATION AND EXPERIMENTAL METHODS}

Single crystals of Cr$_{11}$Ge$_{19}$ were grown by a chemical vapor transport (CVT) method making use of the natural temperature gradient of a tube furnace. About 2.5 g of Cr and Ge powder was sealed in a quartz tube with a molar ratio of 45:55\cite{Han2016}. About 70-100 mg Iodine (~2.2 mg/CC) was used as transfer agent. The temperature was maintained at 880$^{\circ}$C at the deposition zone and 750$^{\circ}$C at the source end. After one-month growth, single crystals of Cr$_{11}$Ge$_{19}$ of average size 1 mm $\times$ 1 mm $\times$ 0.5 mm were obtained. In addition to crystals of Cr$_{11}$Ge$_{19}$, the deposition zone contained single crystals of other phases such as CrGe and pure Ge. Crystals of Cr$_{11}$Ge$_8$ were obtained when the deposition zone was maintained at 1100$^{\circ}$C.

Cr$_{11}$Ge$_{19}$ crystals grown via this method are typically in the shape of a flat-top pyramid as shown in Fig.~\ref{Fig1}(c).  We measured the chemical composition of these crystals via EDX spectra with the composition determined to be Cr$_{38.3\pm2.2}$Ge$_{61.7\pm2.7}$, well within the range expected for a Novotney Chimney ladder compound and in accordance with the previous reports\cite{Ghimire2012,Han2016}. The crystal structure and orientation are determined through single crystal x-ray diffraction. The single crystal refinements for Cr$_{11}$Ge$_{19}$ are summarized in Table.1.

Temperature and field-dependent DC magnetization and AC susceptibility measurements were performed on a Quantum Design (QD) Magnetic Property Measurement System (MPMS). No corrections for demagnetizing fields have been performed. The resisitivity and Hall measurements were carried out in a QD Physical Property measurement system (PPMS). Thin platinum wires were attached to polished surfaces of single crystals via conductive epoxy (Epotek H20E) for charge transport measurements. Resistivity and Hall effect measurements were performed using standard four-terminal low frequency AC techniques using a current of 20 mA at 17 Hz. Field reversal in the Hall measurements was used to compensate for any misalignment of the leads through subtraction of the symmetric part of the field response. The dimension of this crystal was about $1 \times 0.83 \times 0.093 $ mm$^3$ and the applied current densities was 0.21 $\times 10^7 (A/m^2)$.

\begin{table}[ht]
\caption{Single crystal refinements for Cr$_{11}$Ge$_{19}$ at 296(2) K.}
\begin{tabular}{p{0.3\textwidth}cc}
\hline
\hline
 Refined Formula                                 &    Cr$_{11}$Ge$_{19}$           \\ \hline
 F.W.(g/mol)                                     &    1951.21                      \\
 Space group; Z                                  &    P -4n2;4                       \\
 a({\AA})                                        &     5.801 (3)                        \\
 b({\AA})                                        &     5.801 (3)                       \\
 c({\AA})                                        &     52.38 (3)                         \\
 V({\AA}$^3$)                                    &      1762.7 (18)                    \\
 Extinction Coefficient                          &    0.00080 (7)                      \\
 $\theta$ range (deg)                            &      0.777-33.046                  \\
 No. reflections; R$_{int}$                      &     12029; 0.0377                 \\
 No. independent reflections                     &     3218                         \\
 No. parameters                                  &        142                          \\
 R$_1$ : $\omega$ R$_2$ (1$>$2$\delta$(1))       &     0.0531; 0.1389               \\
 Goodness of fit                                 &        1.286                        \\
 Diffraction peak and hole  (e$^-$/{\AA}$^3$)    &      2.073; -3.111                \\
 Absolute structure parameter                    &      0.5 (1)                   \\
 \hline

\end{tabular}
\end{table}

\section{EXPERIMENTAL RESULTS AND DATA ANALYSIS}

\subsection{Magnetization and susceptibility}

\begin{figure}[htb]
\includegraphics[scale=2.5]{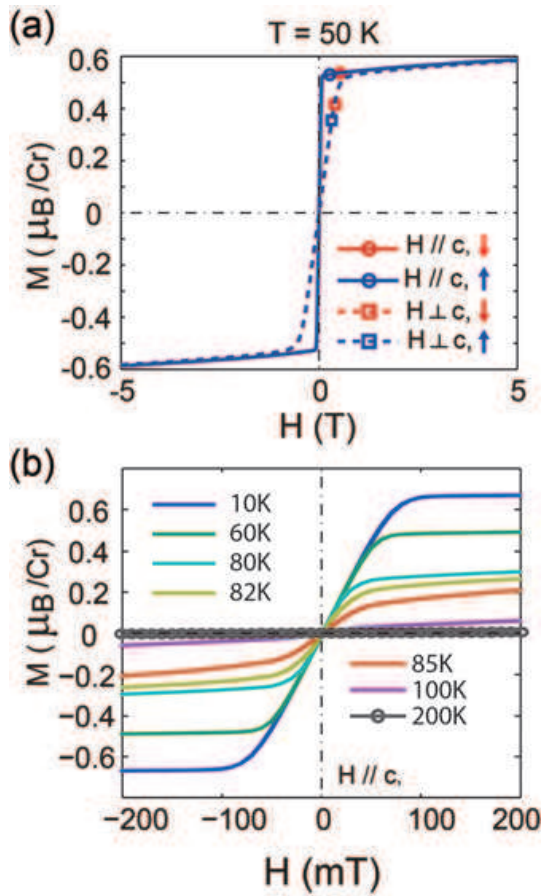}
\caption{ Magnetization. (a) Field-dependence, $H$, of the magnetization, $M$. Solid (dashed) line represents the case where the applied field is along (perpendicular to) the $c$-axis. (b) Magnetization loops measured at a series of temperatures after fielding cooling at 5 T.}
\label{Fig2}
\end{figure}

We first present the magnetic field, $H$, dependence of the magnetization, $M$, with $H$ parallel and
perpendicular to the $c$-axis at 50 K in Fig.~\ref{Fig2} (a). The different saturation fields of $M$ for $H // c$ and $H// ab$ indicate an easy-axis magnetic anisotropy along the $c$-axis. The saturated magnetic moment is small $\sim0.6$ $\mu_B$ at 50 K and 5 T, consistent with previous reports\cite{Takagi2018,Jiang2017,Ghimire2012,Han2016}. In Fig.~\ref{Fig2}(b), we display a series of magnetization loops with $H // c$ at a variety of temperatures after the same field-cooling (FC) process at 5 T. We do not observe any remnant magnetic moment at zero field or a ferromagnetic hysteresis outside of a small anti-hysteresis loop induced by trapped flux in the superconducting magnet\cite{Buchner2018,Sawicki2011} which is too small to be seen in Fig.~\ref{Fig2}(b). The absence of a remnant magnetic moment, along with the easy axis anisotropy, implies that the macroscopic magnetization is completely compensated among regions with different spin orientations which can be either a ferromagnetic ordering with very soft domain-like structure or a periodically modulated ordered state such as a long-period spin density wave or stripe domain order with chiral Bloch domain walls.

\begin{figure}[htb]
\includegraphics[scale=2.5]{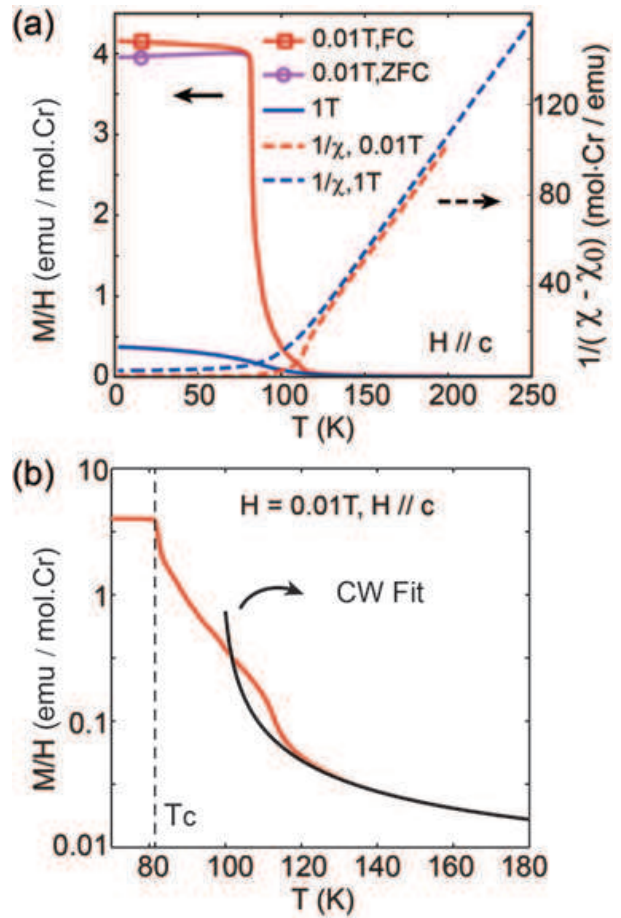}
\caption{(a) DC magnetization, $M/H$ and its inverse, $1/\chi$ at 0.01 and 1 T. (b) $M/H$ at $H = 0.01$ T on the log scale. The black curve is a fit of the Curie-Weiss form to the data. }
\label{Fig3}
\end{figure}

Fig.~\ref{Fig3}(a) displays the DC magnetic susceptibility, $M/H$, along
with the inverse susceptibility $1/\chi = H/M$ for $H = 0.01$ T and 1 T along the $c$-axis. The magnetic
transition temperature $T_c$ is 83 K. We notice that there is a small anomaly where the data deviate significantly from a simple Curie-Weiss (CW) form for $T <$ 110 K as demonstrated in Fig.~\ref{Fig3}(b) where $M/H$ is presented on a logarithmic scale. A fit of a Curie-Weiss formula is illustrated by the solid black curve. The effective magnetic moment determined from the best fit CW form is about 2.7 $\pm$ 0.3 $\mu_B$. A comparison of this effective moment to the saturated magnetic moment found in the data of Fig.~\ref{Fig2} reveals a Rhodes-Wholfarth ratio, $\mu_{eff}/\mu_{sat} \sim 4.5$ revealing a weakly itinerant character. The Weiss temperature, $\Theta_W$, is 100 $\pm$ 2 K, higher than $T_c$, but substantially below the temperature where the anomaly in $M/H$ is observed. In general, the Curie-Weiss law describes the magnetic instability of a thermally disordered spin system at a mean-field level which is expected not to be accurate in proximity to $T_c$ due to the existence of critical fluctuations. For FM materials with critical fluctuations, $M/H$ can diverge at temperatures exceeding $\Theta_W$. However, there are also exceptions such as Sr(Co$_{1-x}$Ni$_x$)$_2$As$_2$\cite{SCNA} where a helical magnetic order is established at a temperature below $\Theta_W$ in the presence of magnetic frustration. We argue that the anomaly at 110 K is unlikely to be due to a second phase as there are no Cr-Ge compounds with a magnetic transition temperature in this range. However, local defects can be induced by the relative sliding between the highly incommensurate Cr- and Ge helices. It is possible that these defects promote a tendency for local enhancement of the magnetization as is observed in magnetic Griffiths phases discovered in disordered magnetic systems\cite{Guo2008,Salamo2002}. The range of ordering temperatures (70 K - 90 K) reported for this compound is likely a result of small differences in stoichiometry between samples grown under different conditions\cite{Han2016,Jiang2017,Ghimire2012} with the level of disorder strongly tied to these differences in stoichiometry.

\begin{figure}[htb]
\includegraphics[scale=1.8]{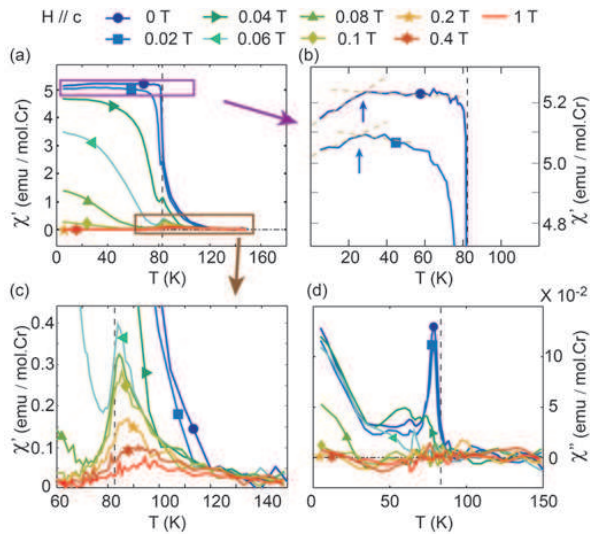}
\caption{ Temperature and field dependence of the AC susceptibility. (a-c) Real part of the AC susceptibility, $\chi'$, in
Cr$_{11}$Ge$_{19}$ at magnetic fields, $H$ identified at the top of the figure. (b) and (c) are the magnification of the two regions enclosed by purple and brown rectangles in panel (a). (d) The corresponding imaginary part of the AC susceptibility, $\chi''$, of Cr$_{11}$Ge$_{19}$. The vertical dashed lines represent $T_c$ at $H$ = 0.}
\label{Fig4}
\end{figure}

To further characterize the magnetic properties of Cr$_{11}$Ge$_{19}$,
we explored the temperature dependence of the real, $\chi'$, and
imaginary, $\chi''$, parts of the AC susceptibility under a series of
magnetic fields parallel to the $c$-axis as shown in Fig.~\ref{Fig4}(a) and
(d). In Fig.~\ref{Fig4}(a), $\chi'$ taken at zero field diverges as the temperature approaches
$T_c$ from above, and exhibits a nearly constant value below $T_c$
suggesting a strongly polarizable state. The principle maximum in
$\chi'$ is suppressed and shifts to lower temperatures as the field
increases while a small peak remains near the zero-field
$T_c$. This is similar to what was observed in ferromagnetic (FM)
AuFe\cite{AuFe} in which the principal maximum is associated with the
motion of domain walls while the small peak around $T_c$ is promoted by the applied
field as shown in Fig.~\ref{Fig4}(c). While the magnitude of this peak is suppressed by the applied magnetic field, it is still observable up to 1 T, separating the nearly polarized FM (NPFM) state and the high-temperature paramagnetic (PM) phase. In Fig.~\ref{Fig4}(b), a small decrease of $\chi'$ is indicated by arrows below 30 K at 0 T and 20 mT. We associate these features with a change in, or stabilization of, the magnetic domains. On the other hand, $\chi''$ in Fig.~\ref{Fig4}(d) displays a strong peak near $T_c$ for $H<40$ mT. An unusual enhancement of $\chi''$ with cooling below $30 $K at low $H$ was also observed, consistent
with a change in the structure and dynamics of magnetic domains in this temperature range. Both of these features disappear
above $0.1$ T, suggesting they are associated with phases located in the low-field region.

\begin{figure}[htb]
\includegraphics[scale=2]{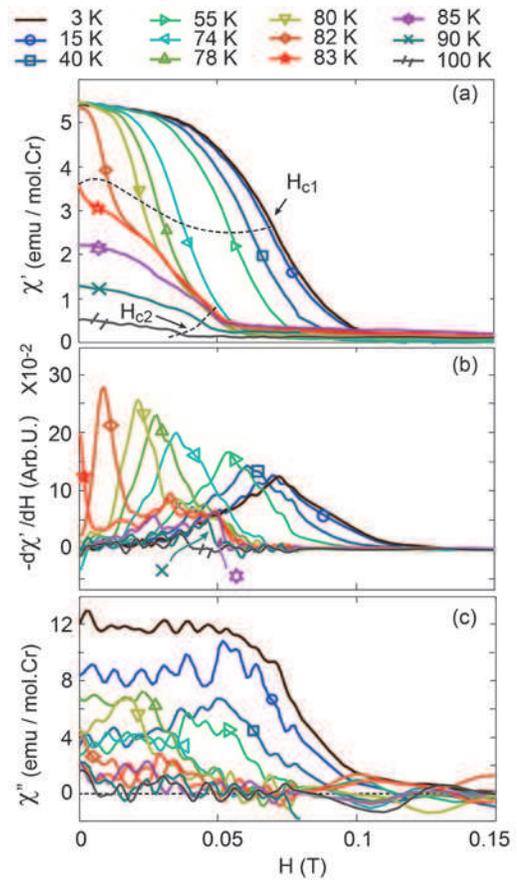}
\caption{ Field dependence of the AC susceptibility. (a) Field dependence of $\chi'$ with the field along the $c$-axis at temperatures, $T$, identified at the top of the figure. (b) The corresponding derivative $-d\chi'/dH$. The maximum defines the critical fields which is labeled as $H_{c1}$ in panel (a). Note that a second peak appears around 0.05 T when the main peak shifts toward zero field as the temperatures increase. The second peak in  $-d\chi'/dH$ leads to the broad feature of $\chi'(H)$ above $T_c$ in panel (a). (c) The imaginary susceptibility $\chi''$ as a function of the magnetic field at the corresponding temperatures.}
\label{Fig5}
\end{figure}

In Fig.~\ref{Fig5}(a) and (c), the field dependence of $\chi'$ and
$\chi''$ are presented at a series of temperatures.  The magnitude of $\chi'$ at zero field is relatively unchanged below $T_c = 83 $K, in accordance with the temperature dependent $\chi'$ in Fig.~\ref{Fig4}(a). As the field increases, $\chi'$ is gradually suppressed to zero in the NPFM state at $H > 0.1 $T and $T < T_c$. We define the critical field $H_{c1}$ by the peaks in -$d\chi'/dH$ as shown in Fig.~\ref{Fig5}(b). From Fig.~\ref{Fig5}(b), it is clear that $H_{c1}$ is reduced to zero as the temperature approaches $T_c$. By contrast, an extra contribution to $\chi'(H)$ appears for $H<0.05 $T at temperatures close to $T_c$. Again, we define $H_{c2}$ as the characteristic field from -$d\chi'/dH$ in Fig.~\ref{Fig5}(b). This additional contribution in $\chi'$ at small fields ($H < H_{c2}$) is responsible for the anomaly observed in the DC susceptibility at $0.01 $T and above $T_c$ [Fig.~\ref{Fig3}(b)].

\begin{figure}[htb]
\includegraphics[scale=1.8]{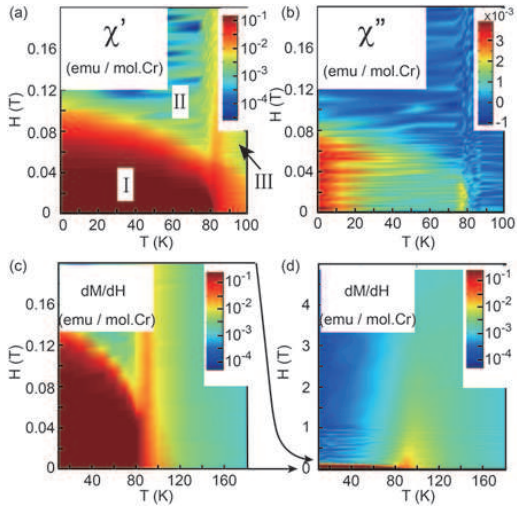}
\caption{AC and DC susceptibility. (a,b) Color plot of the real, $\chi'$ and imaginary, $\chi''$ AC susceptibility as functions of magnetic field, $H$, and temperature, $T$,  from the data in Fig.~\ref{Fig5}. (c,d) The DC susceptibility, $dM/dH$, of Cr$_{11}$Ge$_{19}$ shown in small and large field scales as a comparison.}
\label{Fig6}
\end{figure}

In order to present a complete overview of the magnetic susceptibility  of Cr$_{11}$Ge$_{19}$ as a function of temperature and magnetic field along the $c$-axis, we display a $H-T$ color contour plot of $\chi'$ and $\chi''$ in Fig.~\ref{Fig6}(a) and (b). $\chi'$ is shown on a log scale in Fig.~\ref{Fig6}(a) to highlight features with small magnitude. In Fig.~\ref{Fig6}(a), a red colored ridge with high intensity is located between the NPFM phase and the high-T paramagnetic state and can be tracked back to $T_c$ at zero field. This ridge corresponds to small peaks around $T_c$ in $\chi'(T)$ in Fig.~\ref{Fig4}(c). Moreover, the large values (red area) at small fields appears to continue through the ridge at $T_c$ so that there is substantial intensity above $T_c$. Furthermore, it was found that the imaginary part of the AC susceptibility, $\chi''$, behaves differently above and below 30 K in the low-field region. In Fig.~\ref{Fig6}(c) and (d), the DC susceptibility, $dM/dH$ share the same essential features as $\chi'$. The red-colored ridge at low fields becomes diffuse as the field increases and but is still distinguishable up to 1 T [Fig.~\ref{Fig6}(d)].

\begin{figure}[htb]
\includegraphics[scale=2]{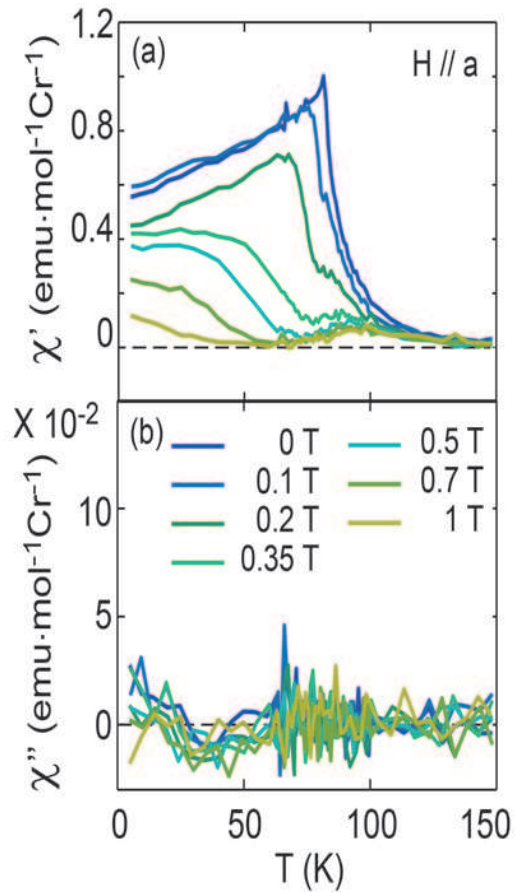}
\caption{ AC susceptibility with field in the $ab$-plane. (a) The temperature dependence of the real part of the AC susceptibility, $\chi'$, under magnetic fields parallel the $a$-axis. (b) The corresponding imaginary part of the AC susceptibility, $\chi''$, as a function of temperature.}
\label{Fig7}
\end{figure}

In addition to the data presented above for fields oriented parallel to the $c$-axis, we have also explored the magnetic response for fields along the $ab$-plane. In Fig.~\ref{Fig7}, we display the temperature dependence of $\chi'_{ab}$ and $\chi''_{ab}$ of Cr$_{11}$Ge$_{19}$ at a series of magnetic fields. $\chi'(T)$ at zero field displays a continuous decrease as the temperature reduced from $T_c$, unlike the plateau seen at zero field in Fig.~\ref{Fig4}(a). The differences seen here and in Fig.~\ref{Fig4}(a) for $H = 0$ are indicative of the intrinsic anisotropy of Cr$_{11}$Ge$_{19}$ as the only difference between these two measurements is the direction of the small AC magnetic fields. Moreover, the evolution of the principal maximum and a secondary peak around $T_c$ is similar to that observed in $\chi'$ with $H$ parallel to $c$. The imaginary part, $\chi''_{ab}$, however, is featureless within the uncertainty of our measurements. In the previous report on polycrystals, a broad peak and a shoulder were observed under magnetic fields and was interpreted as a manifestation of both the itinerant and local moments\cite{Ghimire2012}. With the consideration of both field parallel and perpendicular to the $c$-axis, our measurements on single crystal are consistent with the previous observation but suggest different origins.

\subsection{Charge transport measurements}

\begin{figure}[htb]
\includegraphics[scale=2]{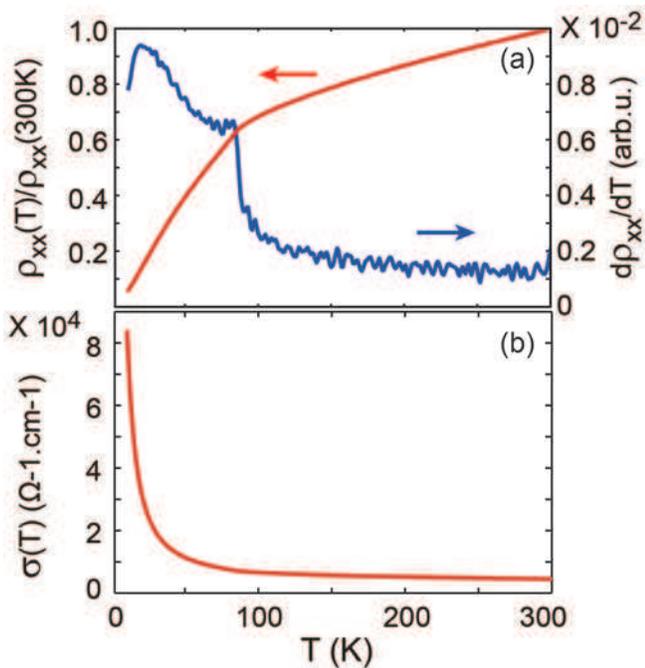}
\caption{Resistivity and Conductivity. (a) The resistivity, $\rho_{xx}$, measured with the current in the $ab$ plane. The blue curve is the derivative $d\rho_{xx}/dT$. (b) The corresponding electric conductivity, $\sigma = 1/\rho_{xx}$ as a function of the temperature.}
\label{Fig8}
\end{figure}

In this section, we will present the results of transport measurements performed on the same single crystal of Cr$_{11}$Ge$_{19}$ with a focus on the topological Hall effect. The resistivity of Cr$_{11}$Ge$_{19}$ with current along the $ab$ plane is measured at zero field as a function of temperature as shown in Fig.~\ref{Fig8}(a). The resistivity is reduced as the temperature decreases, indicating a metallic behavior. The residual-resistivity ratio (RRR) taken as the ration of the resistivity at 300 K to that at 10 K, $\rho_{xx}(300 K)/\rho_{xx}(10 K)$, is about 19 with the resistivity continuing to decline significantly even for temperatures below $10$ K. The derivative, $d\rho_{xx}/dT$, is also plotted as the blue curve in Fig.~\ref{Fig8}(a). A dramatic increase of $d\rho_{xx}/dT$ below the magnetic transition temperature $T_c$ indicates a reduction in the magnetic fluctuations with ordering causing a decrease in the resistivity with cooling. A slight downturn in d$\rho_{xx}$/dT was observed below 20 K, indicating a loss of a mechanism for carrier scattering at low temperatures. In Fig.~\ref{Fig8}(b), we present the electric conductivity, $\sigma$, as $1/\rho_{xx}$. The value of $\sigma$ is less than $10^4$ $(\Omega\cdot cm)^{-1}$ above 50 K and quickly increases below this temperature.

Because the THE is a leading indicator for
the existence of nontrivial spin textures, we measured the topological
Hall resistivity, $\rho^T$, of Cr$_{11}$Ge$_{19}$ through a series of
measurements of $M(H)$, Hall resistivity $\rho_{xy}$ and
magnetoresistance (MR) on the same single crystal with the same field
orientation along the $c$-axis.

\begin{figure}[htb]
\includegraphics[scale=2.5]{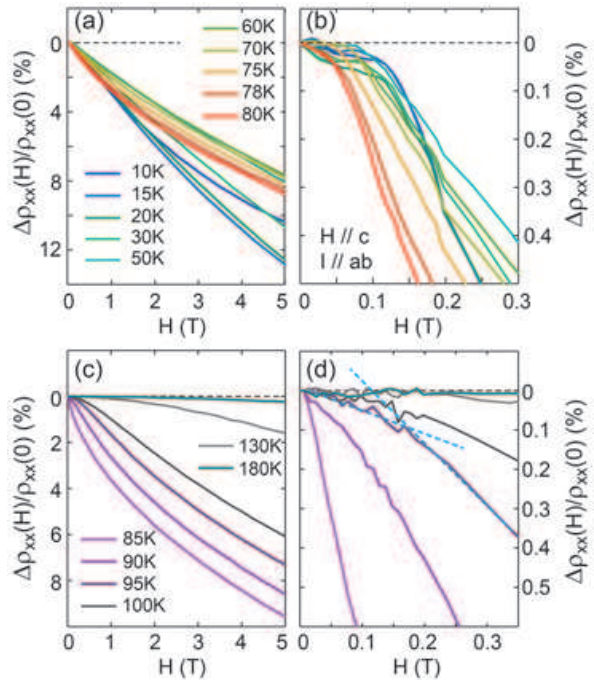}
\caption{ Magnetoresistance, $\Delta\rho_{xx}(H)/\rho_{xx}(0)$ on a large and a small scale to highlight the low field, $H$, regime for temperatures below $T_c$ (a,b) and above $T_c$ (c,d), respectively. The field is applied along the $c$-axis and the current is in the $ab$ plane. $\rho_{xx}$ is defined as the resistivity.}
\label{Fig9}
\end{figure}

In Fig.~\ref{Fig9}, we present the transverse magnetoresistance, $\Delta\rho_{xx}(H)/\rho_{xx}(0) = \frac{\rho_{xx}(H)-\rho_{xx}(0)}{\rho_{xx}(0)}$, with $I \perp c$ and $H // c$ for $T < T_c$ [Fig.~\ref{Fig9}(a) and (b)] and $T > T_c$ [Fig.~\ref{Fig9}(c) and (d)]. A negative MR \cite{NegativeMR} which is characteristic of ferromagnetic materials is observed, reflecting the suppression of electron-spin scattering by the applied magnetic field. We expand the low field region of the magnetoresistance in Fig.~\ref{Fig9}(b) for $T < T_c$ . It is clearly seen that the resistivity is nearly constant below $H_{c1}$ and then quickly drops as the system enters into the NPFM phase. Although there is no magnetic order for $T > T_c$, a convex MR is still observed with a characteristic field illustrated by the dashed lines in Fig.~\ref{Fig9}(d). This characteristic field increases as the temperature increases. At higher magnetic field, a concave behavior of the MR is restored [Fig.~\ref{Fig9}(c)].

\begin{figure}[htb]
\includegraphics[scale=2]{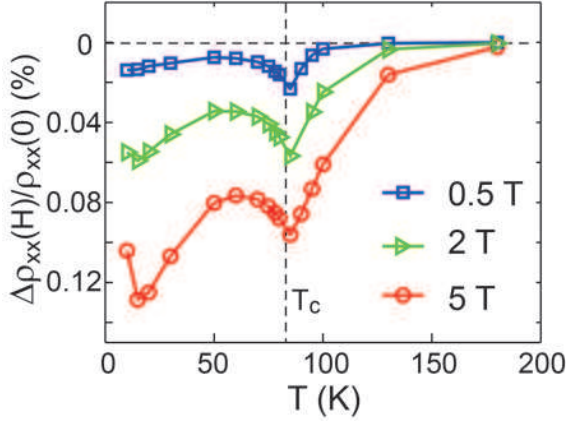}
\caption{Temperature dependence of the magnetoresistance, $\rho_{xx}(H,T)/\rho_{xx}(0,T)$ at magnetic fields $H =$ 0.5 T, 2 T and 5 T. The vertical dashed line represent $T_c$ at zero field.}
\label{Fig10}
\end{figure}

The temperature dependence of the MR is shown in Fig.~\ref{Fig10} in which we plot $\Delta\rho_{xx}(H)/\rho_{xx}(0)$ as a function of temperature at $H$ = 0.5 T, 2 T and 5 T. The local minimum of the MR around $T_c$ indicates a significant contribution of spin fluctuations. We find a significant and unexpected enhancement of the negative MR below 50 K. This effect is particularly significant at high fields as demonstrated by the MR at 5 T in Fig.~\ref{Fig10}. Furthermore, a slight upturn of MR below 15 K is also observed, indicating either a subsidence of the mechanism causing the negative MR or an additional positive contribution at low $T$. These results are consistent with a previous observation\cite{Jiang2017}.

\begin{figure}[htb]
\includegraphics[scale=1.7]{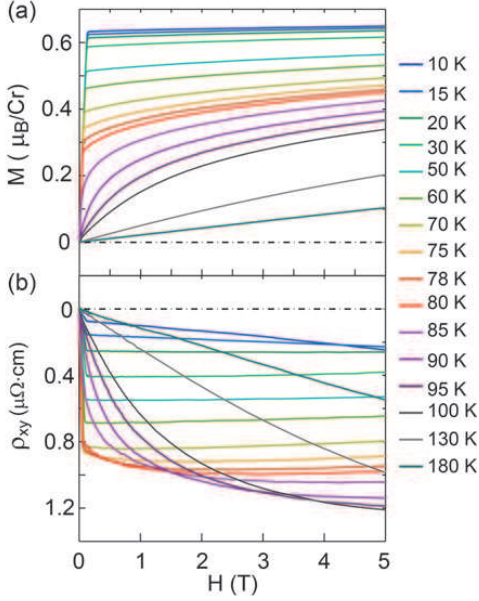}
\caption{ Magnetization, $M$, and Hall resistivity, $\rho_{xy}$, as a function of magnetic field, $H$, at a series of temperatures denoted on the right.}
\label{Fig11}
\end{figure}

\begin{figure}[htb]
\includegraphics[scale=2]{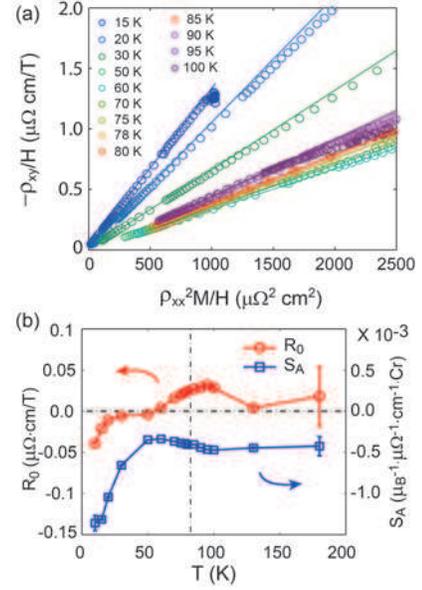}
\caption{ (a)$\frac{\rho_{xy}}{H}$ vs $\frac{\rho^2_{xx}M}{H}$ curves for various temperatures. Solid lines are the fits of a linear form to the data at $H > 1 T$. (b) The normal, $R_0$, and anomalous, $S_A$, Hall coefficients obtained from the fits demonstrated in (a). The error bars represent errors of the fit.}
\label{Fig12}
\end{figure}

In Fig.~\ref{Fig11}, we display $M$ and $\rho_{xy}$ of the same single crystal as a function of the field along the $c$-axis for a series of temperatures. The Hall effect of magnetic materials is commonly written as
\begin{equation}
\rho_{xy} = R_0\textbf{B} + S_A\rho_{xx}^2\textbf{M} + \rho^T.
\end{equation}
in which\cite{Biskyrmion2,Biskyrmion3}, $\rho_{xx}$ is the longitudinal resistivity, $R_0$ and $S_A$
are the normal and anomalous Hall coefficients, and $\textbf{B}$
represents the magnetic flux density. In this analysis we have ignored
contributions from skew scattering to the anomalous Hall resistivity,
which is linear in $\rho_{xx}$, since it is expected to be
insignificant\cite{RMP_AHE} when $\sigma_{xx}$ is smaller than $10^6$
$\Omega^{-1}$cm$^{-1}$, as suggested in Fig.~\ref{Fig8}(b). In order to estimate the coefficients $R_0$ and $S_A$, we plot $\frac{\rho_{xy}}{H}$ vs $\frac{\rho^2_{xx}M}{H}$ in Fig.~\ref{Fig12}(a). In the high field range (H $>$ 1 T) where any contribution from a topological contribution to the Hall effect will likely be very small, the Hall resistivity is expected to obey the standard form of $\frac{\rho_{xy}}{H}$ = R$_0$ + $S_A\frac{\rho^2_{xx}M}{H}$ allowing an accurate determination of $R_0$ and $S_A$. This form is represented in the figure as solid lines. It is clear from the figure that the data are largely described by this form. However, substantial deviation is apparent at lower fields, between 0.2 and 1 T, such that a substantial mismatch was observed between the linear form and the data. The values of $R_0$ and $S_A$ that result from the fitting procedure are shown in Fig.~\ref{Fig12}(b).

\begin{figure}[htb]
\includegraphics[scale=1.8]{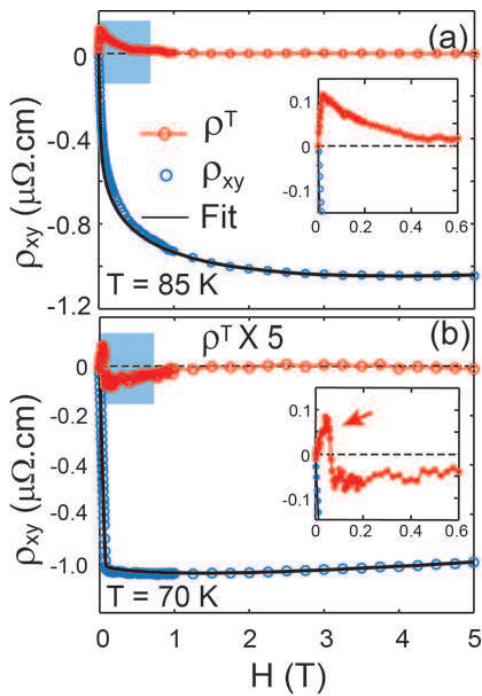}
\caption{ Hall resistivity, $\rho_{xy}$ (blue), and the best fit result of $\frac{\rho_{xy}}{H}$ = R$_0$ + $S_A\frac{\rho^2_{xx}M}{H}$ to these data at high fields. The difference, plotted as red circles, is interpreted as the THE, $\rho^T$. Insets: Magnification of the regions indicated by the blue shading.}
\label{Fig13}
\end{figure}

\begin{figure}[htb]
\includegraphics[scale=1.8]{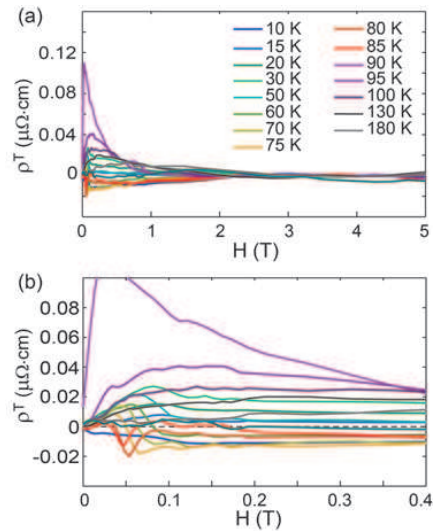}
\caption{ Topological Hall resistivity. (a) The estimated topological Hall resistivity, $\rho^T$ at different temperatures. (b) Magnification of the low fields region in the panel (a). }
\label{Fig14}
\end{figure}

With $R_0$ and $S_A$ determined in this manner, we have calculated the Hall signal expected over the entire range of fields measured using Eq.~(1) without the topological Hall term as presented for two temperatures in Fig.~\ref{Fig13}. The difference is ascribed to the topological Hall term, $\rho^T$. To establish the repeatability of our determination of the topological Hall term, we performed the same sequence of measurements on a second crystal of Cr$_{11}$Ge$_{19}$ having a somewhat different $T_c = 74$ K. The results are largely similar, including reproducing the values of $R_0$ and $S_A$, except that the positive $\rho^T$ at low fields and low temperature ($H < T_{c1}$, $T < T_c$) as denoted by the red arrow in Fig.~\ref{Fig13}(b) is absent in sample 2. We argue that the different thickness (0.18 mm) and current densities (0.06 $\times 10^7 (A/m^2)$) may be responsible for the difference of $\rho^T$ at temperatures below $T_c$ at low fields.

\begin{figure}[htb]
\includegraphics[scale=2]{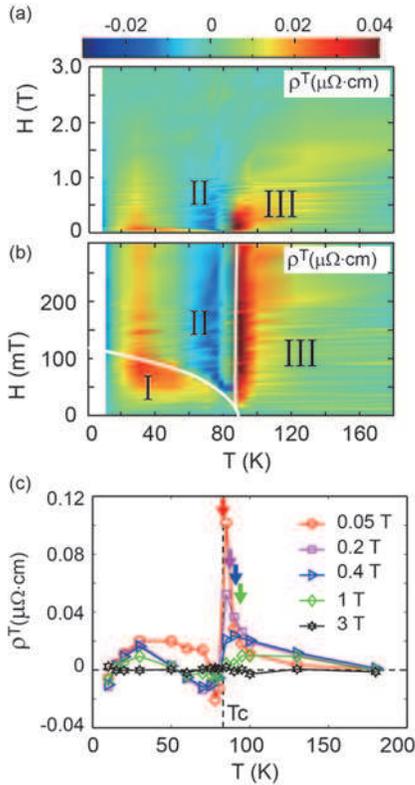}
\caption{ Temperature and Field dependence of the topological Hall resistivity. (a) Magnetic field, $H$, and temperature, $T$, dependence of the topological Hall resistivity,
$\rho^T$. (b) $\rho^T$ shown on a smaller field scale to emphasize the features at low fields. The white curves are the boundaries separating the polarizable low-field, NPFM state and high-Tc PM phases. (c) $T$ dependence of $\rho^T$ at a few selected fields. The arrows represent the locations where $\chi'$ displays a maximum at the corresponding fields. }
\label{Fig15}
\end{figure}

In Fig.~\ref{Fig14}, we plot the resultant $\rho^T$ for sample 1 as a function of field in large (0 $< H <$ 5 T) [Fig.~\ref{Fig14}(a)] and small (0 $< H <$ 0.4 T) [Fig.~\ref{Fig14}(b)] field scales. We find that in spite of a complicated field- and temperature-dependence, the $\rho^T$ display a clear response to the critical field $H_{c1}$ in Fig.~\ref{Fig14}(b), implying a change of the underlying spin texture. To illustrate the evolution of $\rho^T$ with the field and temperature, we present a $H-T$ color contour plot of $\rho^T$ in Fig.~\ref{Fig15}(a). The red (blue) color corresponds to values of $\rho^T$ that are positive (negative). Fig.~\ref{Fig15}(b) displays the same data on a magnified field scale to highlight the low field region. The white lines are the phase boundaries separating the polarizable low-field, NPFM and high-T PM phases. Apparently, $\rho^T$ changes sign near the boundaries of the magnetic phases, as well as displaying a positive value below 30 K where the negative contribution to the MR grows, and the values of $R_0$ and $S_A$ display significant and unexpected temperature dependence. This can be further confirmed in the $T$ dependence of $\rho^T$ as shown in Fig.~\ref{Fig15}(c) in which $\rho^T(T)$ is presented for several fields and where $\rho^T$ is observed to cross the $x$-axis twice in the temperature range 30 K$ < T <$ 100 K for $H$ = 0.05, 0.2, 0.4 and 1 T. Furthermore, the maximum of $\rho^T(T)$ found just above $T_c$ moves towards high temperatures as the field increases, following the evolution of $T_c$ determined by the peak of the AC susceptibility, as indicated by the arrows.

\begin{figure}[htb]
\includegraphics[scale=2]{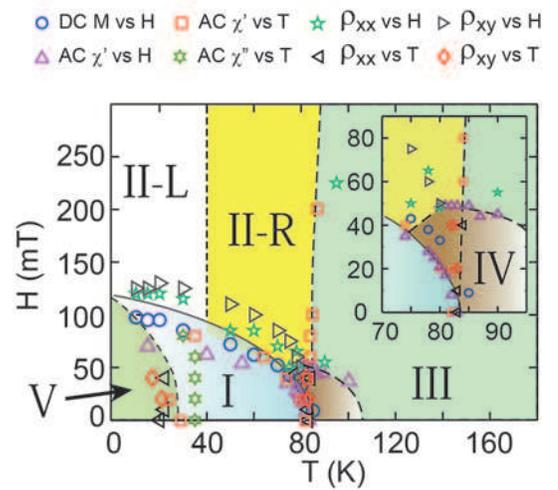}
\caption{ Schematics of the magnetic phase diagram determined
from our magnetic and charge transport measurements. Inset: the magnification
of the area near the critical point. The symbols represent the different measurement techniques listed at the top. }
\label{Fig16}
\end{figure}

In Fig.~\ref{Fig16}, we plot a schematic phase diagram as a summary of our magnetic and charge transport measurements. Region \RomanNumeralCaps{1}, \RomanNumeralCaps{2} and \RomanNumeralCaps{3} are the three main phases referring to the polarizable low-field phase, the NPFM phase,
and the high-T PM states. Region \RomanNumeralCaps{2} is further split into  \RomanNumeralCaps{2}-R and \RomanNumeralCaps{2}-L. In the later, the Hall constants $R_0$, $S_A$ and the MR display significant changes, suggesting that there may be changes to the underlying electronic structure. In region \RomanNumeralCaps{1} below 30 K, the AC magnetic susceptibility is slightly reduced while a large enhancement of $\chi''$ is observed. It remains an open question whether these observations signal a new phase. However, we denote this as region \RomanNumeralCaps{5} to leave open this possibility and to speculate that a magnetic texture may be forming in this region. In addition, we identify an additional phase as region \RomanNumeralCaps{4} just above $T_c$ and below $H_{c2}$ where $\chi'$ is significant.

\section{DISCUSSION AND CONCLUSION}

To place the phase diagram of Fig.~\ref{Fig16} in context, it is helpful to recall previous studies on
two-dimensional dipolar ferromagnets\cite{RMP2000}. In FM films with an easy-axis anisotropy, it is known that the magnetic dipolar interaction drives the
ground state into stripe ordered domains. When a magnetic field is applied, these magnetic stripes break into an intermediate phase of magnetic
bubbles, which ultimately dissolve into the field-polarized
FM state under larger applied fields\cite{Diferro1,Diferro2}. In Cr$_{11}$Ge$_{19}$, the existence of easy-axis spin anisotropy is demonstrated in Fig.~\ref{Fig2}(a). While the stripe domains have not been observed, broken stripes or elongated bubbles are indeed found in LTEM images of thin lamellae at 6 K and zero
field\cite{Takagi2018}. It is plausible that the ground state of Cr$_{11}$Ge$_{19}$ is a stripe order with a zero remnant
magnetization as mentioned above. The idea that the magnetic order becomes more stripe-like at low temperatures corresponds well with the decreased $\chi'$ at low field with a corresponding increase in $\chi''$ below 30 K. With these considerations, we speculate that region \RomanNumeralCaps{5} in Fig.~\ref{Fig16} is related to the stripe-ordered domain phase.

Conversely, a disordered magnetic bubble state may be realized in region \RomanNumeralCaps{1}. As a special kind of magnetic bubbles, bi-skyrmions [Fig.~\ref{Fig1}(d)] are formed from a pair of skyrmions with opposite helicities via attractive interactions. Each skyrmion carries an topological charge (scalar spin chirality) defined as $Q = \frac{1}{4\pi}\int d^2r(\partial_x\textbf{m} \times \partial_y\textbf{m})\cdot\textbf{m}$, where $\textbf{m}$ is a unit vector pointing in the direction of the magnetization. As a spin-polarized electron passes through the spin texture of a skyrmion, it experiences an emergent fictitious magnetic field causing a finite THE. The direction of the fictitious magnetic field is directly related to the sign of $Q$ which distinguishes skyrmions (positive) and antiskyrmions (negative). The THE from bi-skyrmions has been observed as a function of field and temperature in centrosymmetric MnNiGa and MnPdGa systems\cite{Biskyrmion2,Biskyrmion3}, and is also anticipated in Cr$_{11}$Ge$_{19}$. The positive THE that we observe in region \RomanNumeralCaps{1} suggests that bi-skyrmions may be present. However, the temperature range of region \RomanNumeralCaps{1} does not agree with the observation in the LTEM images in which bi-skyrmions were seen at 6 K. We speculate that this may be due to the small thickness of the LTEM sample which allows dipolar fields to stabilize these skyrmion features\cite{Thickness2011}, whereas in our bulk crystalline samples, the effect of these fields is expected to be smaller. Furthermore, since the THE resistivity relies on the fictitious effective field, $B_{eff}$, the local spin polarization of the charge carriers, $P$, and the normal Hall coefficient, $R_0$, we compared these values for both MnNiGa system and Cr$_{11}$Ge$_{19}$, finding comparable values of $\rho^T$. We conclude that there is a distinct possibility of a bi-skyrmion phase in Cr$_{11}$Ge$_{19}$ for sample 1 in the region characterized by a positive $\rho^T$. In addition, the decreased THE in region \RomanNumeralCaps{5} suggests that for bulk samples the magnetic bubble phase transitions to a fully formed magnetic stripe phase at low temperatures. Therefore, we attribute region \RomanNumeralCaps{1} to the bi-skyrmion state in Cr$_{11}$Ge$_{19}$.

Next, we turn our attention to region \RomanNumeralCaps{2}-R at intermediate temperatures and magnetic fields. Across the boundary between region \RomanNumeralCaps{1} and \RomanNumeralCaps{2}-R, we can clearly see a sign change of $\rho^T$ from positive in region \RomanNumeralCaps{1} to negative in region \RomanNumeralCaps{2}-R at temperatures 60 K$ < T <$ 80 K. The sign reversal is sharp and deep for temperatures within 5 K of $T_c$. Sign reversal in $\rho^T$ near $T_c$ was previously reported in MnGe\cite{MnGe2011}, and recently explained as the competition between the THE from the skyrmion lattice and skew scattering from chiral fluctuations\cite{Signchanging}. A similar phenomenon observed here suggests that this mechanism may also apply as we have speculated that bi-skyrmions may be present and where LTEM images indicate bi-skyrmions in thin samples. We note that at 80 K, there are distinct contributions to the negative THE at low and high fields suggesting different origins.

In noncentrosymmetric magnetic materials with spin-orbital coupling, the Dzyaloshinskii-Moriya (DM) interaction in the absence of inversion symmetry is responsible for many interesting magnetic configurations and slow dynamics. In Cr$_{11}$Ge$_{19}$, while the spin dipolar interaction plays a significant role in the formation of bi-skyrmions at low temperatures, it is expected to be weak or even absent in the vicinity of $T_c$ due to the fast spin fluctuations. Near $T_c$, it is anticipated that the anisotropic DM interaction supported by the $D_{2d}$ crystal symmetry may impose a strong effect on the spin dynamics and transport properties. In Fig.~\ref{Fig15}(a)-(c),we observed a large positive $\rho^T$ around $T_c$ which persists deep into the paramagnetic phase in region \RomanNumeralCaps{3}. This strongly suggests that the positive large $\rho^T$ is induced by thermal fluctutaions. Recently, it was proposed that nonzero spin chirality arises as a consequence of the melting of ferromagnetic order by thermal fluctuations in the presence of DM interaction. This mechanism was observed in two different ferromagnetic ultra-thin films of SrRuO$_3$ and V-doped Sb$_2$Te$_3$ where the temperature dependence of $\rho^T$ shows a maximum at $T_c$\cite{ChiralFluctuation}. Our data in Fig.~\ref{Fig15}(c) agrees with this proposition and the peak of $\rho^T$ follows the boundary between the NPFM and PM phases which is derived from the AC susceptibility and marked by the arrows. We note that the evolution betweeen the field polarized and PM phases is a crossover instead of transition under finite magnetic fields.

The application of an external magnetic field explicitly breaks time reversal symmetry of the zero field Hamiltonian. Thus, the transition between FM and PM phases in a FM material under a magnetic field does not involve symmetry breaking. Instead, the order parameter is smeared as a crossover, much like a gas-liquid phase crossover at high pressure rather than a second order phase transition. As the system goes from finite fields to zero field at $T_c$, it enters into the critical region filled by strong fluctuations. In spite of the absence of symmetry breaking, the spin chirality fluctuations in the presence of DM interaction introduce a finite topological charge density and separate the NPFM and PM states. We notice that such spin chirality fluctuations persist up to 1 T above which the NPFM and PM phases are continuously connected.

Furthermore, unlike the isotropic DM interaction in MnSi\cite{Nagaosa2013,DM_DiTusa}, the anisotropic DM interaction in Cr$_{11}$Ge$_{19}$ favors the spin texture of anti-skyrmions which we speculate is the cause of the robustly negative $\rho^T$ that we discover in both samples measured. In Fig.~\ref{Fig15}(a) and (b), the broad blue area with negative THE suggest that it is not an ordinary FM phase and instead contains topologically nontrivial spin textures. Specifically, the opposite sign of THE in region \RomanNumeralCaps{1} and \RomanNumeralCaps{2}-R represents different topological charges and implies that anti-skyrmions may exist in region \RomanNumeralCaps{2}-R for finite fields close to $T_c$ where a negative THE is observed.

The DM interaction relies on the electronic structure\cite{Control_DM} which may evolve with temperature below $T_c$ as suggested by the Hall coefficients (both $R_0$ and $S_A$ show a temperature dependence) in Fig.~\ref{Fig12}(b). Interestingly, we also notice that there is a broad maximum $R_0$ around $T_c$  in Fig.~\ref{Fig12}(b) and this may suggest that the Fermi surfaces and low-energy electronic structure\cite{DOSCr11Ge19} are coupled with the critical fluctuations [Fig.~\ref{Fig10}], which may tune the strength of DM interaction in Cr$_{11}$Ge$_{19}$.
On the other hand, the magnetic dipolar interaction is long-range in nature, and can be screened by fast magnetic fluctuations at temperatures approaching $T_c$. These two factors lead to the dominance of the DM interaction around $T_c$ and a greater importance of dipolar interactions at low temperatures. Therefore, Cr$_{11}$Ge$_{19}$ provides an ideal platform to investigate the competitions among the dipolar interaction, DM interaction and magnetic anisotropy as well as the consequences on the underlying spin textures. Further investigation, such as small angle neutron scattering measurements and a thorough exploration of LTEM images are required to confirm the existence of the purported anti-skyrmion phase in Cr$_{11}$Ge$_{19}$.

Alternatively, the negative THE in region \RomanNumeralCaps{2}-R may also be interpreted as changing character of the dominant charge carriers\cite{Hall3} from electron-like to hole-like, as suggested by the change in sign of $R_0$ in Fig.~\ref{Fig12}(b). However, this can not explain the sign change between region \RomanNumeralCaps{2} and \RomanNumeralCaps{3}. It is also unlikely that bi-skyrmions survive in region \RomanNumeralCaps{2} above $H_{c1}$. Instead, we suggest that Cr$_{11}$Ge$_{19}$ is a nearly compensated metal\cite{Hall1,Hall2} in which the change of sign of $R_0$ can be attributed to a change in the relative scattering rates for electrons and holes.

Finally, the magnetic behavior near the FM critical point in Cr$_{11}$Ge$_{19}$  is very interesting. In Fig.~\ref{Fig16}, Region \RomanNumeralCaps{4} as magnified in the inset corresponds to a region of enhanced magnetic susceptibility and has a boundary determined by $H_{c2}$ and $T_{ano}$. The enhanced magnetic susceptibility in this region is likely not due to simple enhanced critical fluctuations which are responsible for the ridge-like enhancement evident in Fig.~\ref{Fig6}. Instead, region \RomanNumeralCaps{4} is more likely an natural extension of the highly polarizable FM domains in region \RomanNumeralCaps{1}. This suggests that spin clusters which fluctuate substantially in space and time persist well above the melting point, $T_c$, of the magnetic order. The slow dynamics of these spin clusters may account for the temperature dependence of the magnetic susceptibility above $T_c$ at low fields, much like that observed in Griffith's phase systems\cite{Guo2008,Salamo2002}. Recently, the hierarchy of three interactions in MnSi was considered with the weakest being the cubic anisotropy. Here, the hierarchy of interactions leads to an unusual critical regime known as a Brazovskii transition, a fluctuation induced, weakly first order, phase transition\cite{Janoschek2013,Wilson2018}. Cr$_{11}$Ge$_{19}$ clearly does not fit this description because of the very different crystal symmetry. Instead, region IV in the phase diagram of Fig.~\ref{Fig16} may be the result of a more complex critical regime that reflects the complex interactions present in this system. These interactions include the increasing importance of the uniaxial anisotropy with cooling and an antisymmetric DM interaction which prefers an alternating chirality.

In summary, we have carried out a series of magnetic and charge transport
measurements on single crystals of Cr$_{11}$Ge$_{19}$ unveiling a rich
phase diagram. A set of interesting phases is postulated from the
results of measurements of the AC susceptibility and the THE adding to the recently discovered bi-skyrmion phase in this material. A second topological non-trivial phase is postulated at temperatures approaching $T_c$ which we believe may be
an anti-skyrmion phase consistent with the crystalline
symmetry. In addition, the magnetic susceptibility at low fields is
significantly enhanced above $T_c$, implying a cluster or disordered phase likely
 due to the anisotropy, crystalline disorder, and the DM interaction, in contrast with that found in MnSi from the weak cubic anisotropy.
 Considering the likely difference in the temperature dependencies of the DM interaction and
the magnetic dipolar interaction, we argue that a transition between
the bi-skyrmion state at low temperature and an anti-skyrmion state near $T_c$ may be realized in Cr$_{11}$Ge$_{19}$. This is reminiscent of LTEM images of Mn$_{1.4}$Pt$_{0.9}$Pd$_{0.1}$Sn\cite{transition1,transition2}, correlating the sign change of the THE with the change from bi-skyrmion to anti-skyrmion phase.

\section{Acknowledgments}

Y.L. is grateful to D.L. Gong, L.Y. Xing, and J.H. Chen for the
assistance on measurements and thank R.Y. Jin and Z.T. Wang for
helpful discussion. We acknowledge D.M. Cao, Y. Mu, X.C. Wu at the
Shared Instrumentation Facility (SIF), LSU for chemical analysis. This
project is supported by the U.S. Department of Energy under EPSCoR
Grant No. DESC0012432 with additional support from the Louisiana Board
of Regents.

{}


\begin{thebibliography}{}
\bibitem{Skyrmion1} S. M\"{u}hlbauer, B. Binz, F. Jonietz, C. Pfleiderer, A. Rosch, A. Neubauer, R. Georgii, and P. B\"{o}ni, Science, {\bf 323}, 915-919 (2009).
\bibitem{Skyrmion2} Stefan Heinze, Kirsten Von Bergmann, Matthias Menzel, Jens Brede, Andr\'{e} Kubetzka, Roland Wiesendanger, Gustav Bihlmayer, and Stefan Bl\"{u}gel, Nat. Phys. {\bf 7}, 713-718 (2011).
\bibitem{Skyrmion3} S. Seki, X.Z. Yu, S. Ishiwata, and Y. Tokura, Science, {\bf 336}, 198-201 (2012).
\bibitem{Dynamics} Xichao Zhang, Jing Xia, Yan Zhou, Xiaoxi Liu, Han Zhang, and Motohiko Ezawa, Nat. commun. {\bf 8}, 1717 (2017).
\bibitem{Nagaosa2013} Naoto Nagaosa, and Yoshinori Tokura, Nat. Nanotechnol. {\bf 8}, 899-911 (2013).
\bibitem{DM_DiTusa} C. Dhital, L. DeBeer-Schmitt, Q. Zhang, W. Xie, D.P. Young, and J.F. DiTusa, Phys. Rev. B {\bf 96}, 214425 (2017).
\bibitem{Nayak2017} Ajaya K. Nayak, Vivek Kumar, Tianping Ma, Peter Werner, Eckhard Pippel, Roshnee Sahoo, Franoise Damay, Ulrich K. R\"{o}{\ss}ler, Claudia Felser£¬ and Stuart S.P. Parken, Nature {\bf 548}, 561-566 (2017).
\bibitem{Huang2017} Siying Huang, Chao Zhou, Gong Chen, Hongyi Shen, Andreas K. Schmid, Kai Liu, and Yizheng Wu, Phys. Rev. B {\bf 96}, 144412 (2017).
\bibitem{Biskyrmion1} Licong Peng, Ying Zhang, Wenhong Wang, Min He, Lailai Li, Bei Ding, Jianqi Li, Young Sun, X.-G. Zhang, Jianwang Cai, Shouguo Wang, Guangheng Wu, and Baogen Shen, Nano Lett. {\bf 17}, 7075-7079 (2017).
\bibitem{Biskyrmion2} Wenhong Wang, Ying Zhang, Guizhou Xu, Licong Peng, Bei Ding, Yue Wang, Zhipeng Hou, Xiaoming Zhang, Xiyang Li, Enke Liu, Shouguo Wang, Jianwang Cai, Fangwei Wang, Jianqi Li, Fengxia Hu, Guangheng Wu, Baogen Shen, and Xi-Xiang Zhang, Adv. Mater. {\bf 32}, 6887-93 (2016).
\bibitem{Biskyrmion3} Xiaofei Xiao, Licong Peng, Xinguo Zhao, Ying Zhang, Yingying Dai, Jie Guo, Min Tong, Ji Li, Bing Li, Wei Liu, Jianwang Cai, Baogen Shen, and Zhidong Zhang, Appl. Phys. Lett. {\bf 114}, 142404 (2019).
\bibitem{Biskyrmion4} Xiyang Li, Shilei Zhang, Hang Li, Diego Alba Venero, Jonathan S White, Robert Cubitt, Qingzhen Huang, Jie Chen, Lunhua He, Gerrit van der Laan, Wenhong Wang, Thorsten Hesjedal, and Fangwei Wang, Adv. Mater. {\bf 31}, 1900264 (2019).
\bibitem{transition1} Licong Peng, Rina Takagi, Wataru Koshibae, Kiyou Shibata, Kiyomi Nakajima, Taka-hisa Arima, Naoto Nagaosa, Shinichiro Seki, Xiuzhen Yu, and Yoshinori Tokura, Nat. Nanotechnol. {\bf 15}, 181-186 (2020).
\bibitem{transition2} Jagannath Jena, B\"{o}rge G\"{o}bel, Tianping Ma, Vivek Kumar, Rana Saha, Ingrid Mertig, Claudia Felser, and Stuart S.P. Parkin, Nat. Commun. {\bf 11}, 1115 (2020).
\bibitem{Takagi2018} R. Takagi, X.Z. Yu, J.S. White, K. Shibata, Y. Kaneko, G. Tatara, H.M. R{\o}nnow, Y. Tokura, and S. Seki, Phys. Rev. Lett. {\bf 120}, 037203 (2018).
\bibitem{NCladder} Daniel C. Fredrickson, Stephen Lee, and Roald Hoffmann, Inorg. Chem. {\bf 43}, 6159-6167 (2004).
\bibitem{Mn3Ge5} Rodrigo Castillo, Walter Schnelle, Matej Bobnar, Raul Cardoso-Gil, Ulrich Schwarz and Yuri Grin, Zeitschrift f\"{u}r anorganische und allgemeine Chemie, {\bf 646}, 256-262 (2020).
\bibitem{Han2016} Hui Han, Lei Zhang, Xiangde Zhu, Haifeng Du, Min Ge, Langsheng Ling, Li Pi, Changjin Zhang, and Yuheng Zhang, Sci. Rep. {\bf 6}, 39338 (2016).
\bibitem{Ghimire2012} N.J. Ghimire, M.A. McGuire, D.S. Parker, B.C. Sales, J.-Q. Yan, V. Keppens, M. Koehler, R.M. Latture, and D. Mandrus, Phys. Rev. B {\bf 85}, 224405 (2012).
\bibitem{Jiang2017} N. Jiang, Y. Nii, R. Ishii, Z. Hiroi, and Y. Onose, Phys. Rev. B {\bf 96}, 144435 (2017).
\bibitem{Buchner2018} M. Buchner, K. H\"{o}fler, B. Henne, V. Ney, and A. Ney, J. Appl. Phys. {\bf 124}, 161101 (2018).
\bibitem{Sawicki2011} M. Sawicki, W. Stefanowicz, and A. Ney, Semicond. Sci. Technol. {\bf 26}, 064006 (2011).
\bibitem{SCNA} Yu Li, Zhonghao Liu, Zhuang Xu, Yu Song, Yaobo Huang, Dawei Shen, Ni Ma, Ang Li, Songxue Chi, Matthias Frontzek, Huibo Cao, Qingzhen Huang, Weiyi Wang, Yaofeng Xie, Rui Zhang, Yan Rong, William A. Shelton, David Young, J.F. DiTusa, and Pengcheng Dai, Phys. Rev. B {\bf 100}, 094446 (2019).
\bibitem{Guo2008} S. Guo, D. P. Young, R. T. Macaluso, D. A. Browne, N. L. Henderson, J. Y. Chan,
L. L. Henry, and J. F. DiTusa, Phys. Rev. Lett. {\bf100}, 017209 (2008).
\bibitem{Salamo2002} M.B. Salamon, P. Lin, and S.H. Chun, Phys. Rev. Lett. {\bf 88}, 197203 (2002).
\bibitem{AuFe} I. Maartense and Gwyn Williams, Phys. Rev. B {\bf 17}, 377 (1978).
\bibitem{NegativeMR} Hiroshi Yamada and Satoshi Takada, Prog. Theor. Phys. {\bf 48}(6), 1828-1848 (1972).
\bibitem{RMP_AHE} Naoto Nagaosa, Jairo Sinova, Shigeki Onoda, A.H. MacDonald, and N.P. Ong, Rev. Mod. Phys. {\bf 82}, 1539 (2010).
\bibitem{RMP2000} K. De'bell, A.B. Maclsaac, J.P. Whitehead, Rev. Mod. Phys. {\bf 72}, 225 (2000).
\bibitem{Diferro1} Rogelio D\'{\i}az-M\'{e}ndez, and Roberto Mulet, Phys. Rev. B {\bf 81}, 184420 (2010).
\bibitem{Diferro2} T. Garel and S. Doniach, Phys. Rev. B {\bf 26}, 325 (1982).
\bibitem{Thickness2011} X.Z. Yu, N. Kanazawa, Y. Onose, K. Kimoto, W.Z. Zhang, S. Ishiwata, Y. Matsui, and Y. Tokura, Nat. Mater. {\bf 10}, 106-109 (2011).
\bibitem{MnGe2011} N. Kanazawa, Y. Onose, T. Arima, D. Okuyama, K. Ohoyama, S. Wakimoto, K. Kakurai, S. Ishiwata, and Y. Tokura, Phys. Rev. Lett. {\bf 106}, 156603 (2011).
\bibitem{Signchanging} Hiroaki Ishizuka, and Naoto Nagaosa, Sci. Adv. {\bf 4},eaap9962 (2018).
\bibitem{ChiralFluctuation} Wenbo Wang, Matthew W. Daniels, Zhaoliang Liao, Yifan Zhao, Jun Wang, Gertjan Koster, Guus Rijinders, Cui-Zu Chang, Di Xiao and Weida Wu, Nat. Mater. {\bf 18}, 1054-1059 (2019).
\bibitem{Control_DM} Takashi Koretsune, Naoto Nagaosa, and Ryotaro Arita, Sci. Rep. {\bf 5}, 13302 (2015).
\bibitem{DOSCr11Ge19} P. P\'{e}cheur, G. Toussaint, H. Kenzari, B. Malaman, and R. Welter, J. Alloy. Compd. {\bf 262-263}, 363 (1997).
\bibitem{Hall3} David LeBoeuf, Nicolas Doiron-Leyraud, Julien Levallois, R. Daou, J.-B. Bonnemaison, N.E. Hussey, L. Balicas, B.J. Ramshaw, Ruixing Liang, D.A. Bonn, W.N. Hardy, S. Adachi, Cyril Proust, and Louis Taillefer, Nature {\bf 450}, 533-536 (2007).
\bibitem{Hall1} N.W. Ashcroft, Phys. Kondens. Materie {\bf 9}, 45-53 (1969).
\bibitem{Hall2} Manuel Zingl, Jernej Mravlje, Markus Aichhorn, Olivier Parcollet, and Antoine Georges, npj Quantum Mater. {\bf 4}, 35 (2019).

\bibitem{Janoschek2013} M. Janoschek, M. Garst, A. Bauer, P. Krautscheid, R. Georgii, P. B\"{o}ni, and C. Pfleiderer, Phys. Rev. B {\bf 87}, 134407 (2013).
\bibitem{Wilson2018} J. D. Bocarsly, R. F. Need, Ram Seshadri, and S. D. Wilson, Phys. Rev. B {\bf 97}, 100404(R) (2018).


\end{thebibliography}
\end{document}